\documentclass[twocolumn,floatfix,showpacs,preprintnumbers,amsmath,amssymb,showkeys,prb]{revtex4-1} 

\usepackage[T1]{fontenc}
\usepackage[english]{babel}
\usepackage{graphicx}
\usepackage{amsmath,amssymb,mathrsfs}
\usepackage{hyperref}
\usepackage{caption}

\begin{document}

\title{Thermodynamics of the $\alpha$-$\gamma$ cerium phase transition from first principles}

\author{J. Bieder and B. Amadon}

\affiliation{CEA, DAM, DIF, F 91297 Arpajon, France} 

\begin{abstract} 
We present a thermodynamical investigation of the $\alpha\rightleftharpoons\gamma$ transition of Ce using Density Functional Theory within the projector augmented wave framework combined with the Dynamical Mean Field Theory.
First, we confirm that without spin-orbit coupling, no transition appears at zero temperature. 
Secondly, we extend the same conclusion to finite temperature with a slight difference: a crossover is observed both in temperature and pressure between the $\alpha$ and $\gamma$ phases. 
This is obviously visible as a softening of the bulk modulus. 
Thirdly, we show the leading role of entropy for the description of the equation of state of cerium. 
Lastly, we discuss the role of spin-coupling, and we argue that neglecting the spin-orbit coupling is roughly equivalent to a renormalization of temperature. 
Indeed at 800 K, both our variation of thermodynamical quantities and our spectral functions describe the experimental data.
\end{abstract}


\maketitle

\section{Introduction}
Cerium, a rare earth metal, undergoes an isomorphic $\alpha$, $\gamma$ first order solid-solid transition with a volume ($V$) collapse of about 14\% \cite{Koskimaki1978,Lipp2012,Decremps2011}. 
The corresponding transition line ends at a critical point (CP) around 1.5 GPa and 480 K \cite{Lipp2012,Decremps2011}. 
The (larger volume) $\gamma$ phase exhibits a Curie-Weiss behavior for the magnetic susceptibility and is Pauli-like in the (smaller volume) $\alpha$ phase. 
This is interpreted as $4f$ electrons being localized in the $\gamma$ phase, giving rise to local moments and contributing weakly to the electronic bonding whereas in the $\alpha$ phase, the $4f$ electrons participate in both the bonding and the formation of quasi-particles. 
The $\alpha\rightleftharpoons\gamma$ phase transition of Ce is hence a model system for volume collapse phase transition due to the delocalization of localized electrons under increase of pressure or decrease of temperature $(T)$.
A first principles description would then have a huge impact on the description of other correlated systems, although it remains a challenging study.

The transition was first described \cite{Schuch1950} by the promotion of a $f$ electron into $spd$ orbitals under pressure, but it was refuted by experiments \cite{Gustafson1969,Kornstadt1980}.
Several models have thus been built.
The Mott transition (MT) model \cite{Johansson1974} assumes that the transition is driven by the overlap of $f$ orbitals while the Kondo Volume Collapse (KVC) \cite{Allen1982,Lavagna1982} model assumes that the hybridization of $f$ orbitals with $spd$ plays the main role.
Both are qualitatively coherent with the localization delocalization picture.
As a consequence, an {\it ab-initio} description which includes all hybridization effects is needed.

From an {\it ab-initio} point of view, Local (LDA) and Gradient (GGA) functionals for Density Functional Theory (DFT) roughly describe the $\alpha$ phase and its delocalized electron.
DFT+U \cite{Shick2001} or Self-Interaction Correction \cite{Johansson1995,Luders2005} are nevertheless required to describe the localization of the $f$ electrons in the $\gamma$ phase.
Thereby, first attempts to describe the transition from an {\it ab-initio} framework used the assumption of a pseudo alloy, especially for the finite temperature extension \cite{Svane1994,Szotek1994,Johansson1995,Luders2005}.

In the recent past, the exact exchange (EXX) functional with the Random Phase Approximation (RPA) correlation was able to describe qualitatively the transition at 0 K \cite{Casadei2012}.
It was, however, only with the advent of pioneering calculations using the combination of Density Functional Theory with Dynamical Mean Field Theory (DFT+DMFT) that spectral features were at least qualitatively described from {\it ab-initio} \cite{Zolfl2001,Held2001}, as well as optical spectra \cite{Haule2005} and magnetic properties \cite{Streltsov2012}.
Concerning the thermodynamics of the transition at finite temperature, the calculation of internal energy as a function of volume was computed in DFT+DMFT \cite{Held2001,McMahan2003,Amadon2006}.
Nonetheless, the precision was limited by the Atomic Sphere Approximation (ASA), the lack of self-consistency over charge density, the noise of the  Hirsch Fye Quantum Monte Carlo (QMC) \cite{Hirsch1986} algorithm because of limited computational power.
That being so, the question of the appearance of a Maxwell's common tangent on the internal energy as a function of the volume at finite temperature is still opened mainly because of the non negligible error bars \cite{Held2001,McMahan2003,Amadon2006}.
Recently, \citet{Lanata2013} used the Gutzwiller approximation restricted to zero temperature to emphasize the existence of a phase transition when spin-orbit coupling (\textsc{soc}) is taken into account.
These calculations thus contest the existence of another critical point at the other side of the transition line when extrapolated to zero temperature as considered experimentally on alloys \cite{Lashley2006} and theoretically on model systems \cite{deMedici2005}. 
Regarding finite temperature, several recent works were successfully carried out to describe  temperature phase transitions in correlated systems such as iron~\cite{Leonov2011} or V$_2$O$_3$~\cite{Grieger2012}. 
Even so, the calculation of the complete free energy was not carried out.
A recent work in iron~\cite{Pourovskii2013} however, applied successfully a coupling constant integration to compute entropy of three different phases.
For cerium however, beyond pioneering works \cite{McMahan2003}, an accurate calculation of free energy along the transition is still lacking.

In this article, we first review in section~\ref{sec:details} the framework and scheme with the convergence parameters used.
We then examine in section~\ref{sec:sc} the need for a full self-consistent DFT+DMFT scheme before reporting our extensive calculations of the $\alpha-\gamma$ transition in cerium in section~\ref{sec:entropy}.  
In this section we analyze the entropic effect at low and high temperature and derivate the pressure from the free energy.
First, we confirm the picture of \citet{Lanata2013} who used a Gutzwiller scheme: without spin-orbit coupling, no transition appears at zero temperature. 
This is important because the DFT+Gutzwiller scheme can be viewed as a further approximation to DFT+DMFT.
Second, we extend this conclusion to finite temperature. 
Above zero temperature, however, a crossover is observed between the $\alpha$ and $\gamma$ phases. 
The bulk modulus which is the first derivative of pressure as a function of volume, has a softening around the transitional volume. 
Third, we show the leading role of entropy for the description of the equation of state of cerium. 
In section~\ref{sec:soc}, we finally examine the role of spin-orbit coupling and we argue that neglecting it is roughly equivalent to a renormalization of temperature. 
Both the variations of thermodynamical quantities and spectral function at 800 K validate this hypothesis.

\begin{figure}[btp]
  \centering
  \includegraphics{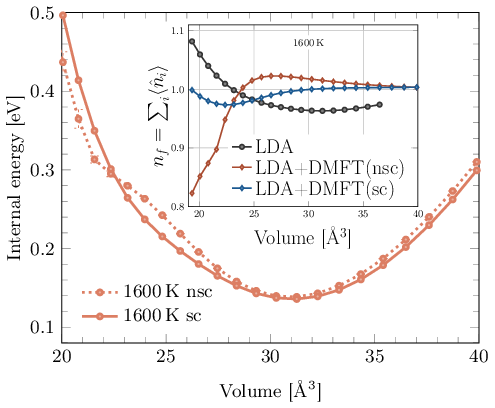}
  \caption{(color online) Two proofs for the need of self consistency over density in LDA+DMFT calculations for cerium.
    Internal energy at 1600 K : both calculations are notably different for small and large volumes.
    Inset : total number of $4f$ electrons at 1600 K ($n_f$) in LDA, LDA+DMFT (nsc) and LDA+DMFT (sc). 
    The huge difference of this number of $4f$ electrons produces a significant modification of the total electronic density and thus changes the internal energy.
  } 
  \label{fig:scf} 
\end{figure}
\section{Computational details\label{sec:details}}
We used a recent implementation of DFT+DMFT in the ABINIT code \cite{Gonze2009}, with an accurate PAW \cite{Blochl1994} basis \cite{Torrent2008}, the self-consistency over density \cite{Amadon2012} and an efficient implementation \cite{Bieder2013} of a continuous time QMC solver with the hybridization expansion \cite{Werner2006} (CT-Hyb). 
The same atomic data as in Ref. \cite{Amadon2012} were used. 
In particular, $5s$ and $5p$ semi-core states are included in the valence.
We used a $10\times10\times10$ $k$-mesh grid and a cut-off of $30$ Ha for the plane waves.
The local orbitals were Wannier functions as in Ref. \cite{Amadon2012}, following the scheme of Ref. \cite{Amadon2008}.
We used a recent self-consistent cRPA implementation for the calculation of $U$ \cite{Amadon2014} in our Wannier basis. 
We found the screened coulomb interaction to be $U=5.9$ eV for the $\gamma$ phase. 
We thus decide to use a screened coulomb interaction of 6 eV as used also in previous works \cite{Held2001,Amadon2006, Streltsov2012} and we neglect the variation of $U$ as a function of volume.
DFT+DMFT calculations are performed until convergences of local Green's function and electronic density. 
$10^{10}$ to $5\cdot10^{11}$ steps are performed for each CT-Hyb run according to the temperature, so that the stochastic noise over internal energy is less than 0.2 meV (smaller than the mark size).
The zero energy (for internal and free energies) is arbitrary chosen as we are only interested in variations and differences.
As those calculations are computationally expensive, we did not take into account the spin-orbit coupling for the internal and free energy curves, although we did include the spin-orbit coupling for the spectral functions in the last part of this study.
The internal energy $E$ in DFT+DMFT can be computed from the DMFT density $\rho_{\rm DMFT}$ and impurity Green's function \cite{Amadon2006} for given value sets of screened coulomb interaction $U=6$ eV, volume $V$, and temperature $T$. 
We used for the double counting energy $E_\text{DC}$ the full localized limit as in previous works \cite{Held2001,Amadon2006,Streltsov2012}.

\section{Importance of full self-consistency for the internal energy\label{sec:sc}}
Before exploring the physics of cerium from a first principle (with the same treatment for both phases and no adjustment) point of view, one has to be careful about the scheme used.
The main cost in the DFT+DMFT scheme is the charge self-consistency that implies extensive cpu time due to the large number of impurity problems that have to be solved.
In the previous works, the authors usually chose either self-consistency (sc) over electronic density (e.g. \cite{Amadon2012}) with a not accurate impurity solver, either non self-consistent (nsc) calculations with an accurate solver (e.g. \cite{Streltsov2012}).
Sometimes the solver precision was weak because of the stochastic noise in combination of a non self-consistent calculation so there was a remaining imprecision\cite{Held2001,McMahan2003,Amadon2006}.
That is why we present our accurate (involving PAW calculations and CT-Hyb resolution of the QMC) LDA+DMFT internal energy calculations with and without charge consistency.
From the DMFT density $\rho_{\rm DMFT}$, the internal energy $E$ in DFT+DMFT can be computed \cite{Amadon2006} for given values of screened coulomb interaction $U=6$ eV, volume $V$, and temperature $T$.
\begin{equation}
  E(U) = E_{LDA}(\rho_{\rm DMFT}) - \sum_\lambda \epsilon_\lambda^{\text{KS}} + \langle H_{\text{KS}} \rangle + E_{\rm int},
  \label{eq:energy}
\end{equation}
where $E_{\rm int}=\langle H_U \rangle - E_\text{DC}$. $E_{\rm int}$ is the only part of the energy which has an explicit dependence of $U$.
$\langle H_U \rangle$ is computed inside the CT-Hyb \cite{Gull2011} and we use for the double counting energy $E_\text{DC}$ the full localized limit double counting as in previous works \cite{Held2001,Amadon2006,Streltsov2012}.
On Fig.~\ref{fig:scf}, the internal energy of cerium is plotted with respect to the volume at a temperature of 1600 K. 
We notice for both large and small volumes the energy difference between LDA+DMFT(nsc) and LDA+DMFT(sc) is around 20 meV which is the same order as the experimental difference in internal, free energies and entropy between $\alpha$ and $\gamma$-Ce \cite{Amadon2006,Decremps2011}.
This variation can be explained by the difference of the number of $4f$ electrons $n_f$ as depicted by the inset on Fig.~\ref{fig:scf}.
Indeed, the difference of $n_f$ between the converged LDA calculation and the first converged DMFT loop is major, mainly for small volumes.
We argue that at the end of this first DMFT loop, the $n_f$ electrons leave the $f$ orbitals to join the lower $spd$ orbitals.
That might cause a large change in the LDA density that as to be corrected with the self-consistency.
When the LDA+DMFT calculation is converged, then $n_f$ is smoother and is between LDA and LDA+DMFT(nsc) results.
This small valley can be interpreted by the increase of the hybridization of $4f$ electrons with the $spd$ which is due to an increase of the overlap as suggested by \cite{McMahan2003}.
For smaller volumes, the pressure brings $spd$ electrons in the $f$ orbitals.
There we conclude that the electronic density has to be converged within LDA+DMFT to study cerium.
This was highlighted before by~\citet{Lanata2013} in the context of the Gutzwiller approximation.
Here, we confirm this conclusion using the DFT+DMFT framework.

Our fully consistent calculations of the internal energy for several volumes and temperatures are presented on the middle panel of Fig.~\ref{fig:F_V} (dashed light lines).
Our data show the unambiguous existence of two inflection points and a negative curvature in the energy versus volume curves.
It originates from the Kondo stabilization as discussed in earlier works \cite{McMahan2003,Amadon2006}.
At high temperature, the Kondo effect lowers the internal energy only at small volumes (at 800K, only below $\simeq$ 30 \AA$^3$) so that a negative curvature clearly appears at intermediate volume (at 800K, around 30 \AA$^3$).
At lower temperature, it stabilizes a larger range of volumes and thus the negative curvature is less visible  (it extends over the whole range of the transitional volume).
This negative curvature was not unambiguously visible in earlier studies due to the large statistical noise and lack of precision \cite{McMahan2003,Amadon2006}.
We note that the difference of internal energy between alpha and gamma cerium as a function of temperature decreases and this is coherent with the increase of $\Delta E$ as plotted in Fig. 1 of Refs. \onlinecite{Amadon2006} and \onlinecite{Decremps2011}. 

\section{Entropic effect on thermodynamical quantities\label{sec:entropy}}
\begin{figure}[tbp]
  \centering
  \includegraphics{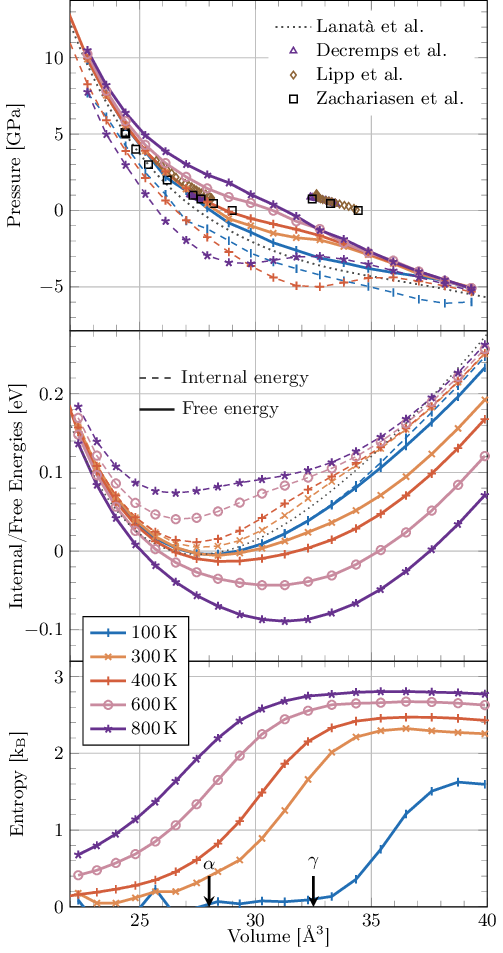}
  \caption{(color online) Upper panel : Clapeyron diagram with our theoretical results with (solids lines) and without (dashed lines) the entropic contribution, Gutzwiller results (dotted lines) from~\citet{Lanata2013} and experimental data of~\citet{Lipp2008} at 293 K (diamonds),~\citet{Decremps2011} at 334 K and~\citet{Zachariasen1977} at 300 K (squares). Middle panel: internal (dashed lines) and free (solid lines) energies versus volume computed in LDA+DMFT(CT-Hyb) for different temperatures.
    Lower panel: entropic contribution for the same temperatures versus volume.
    Arrows indicate the experimental volumes of each phase at 400 K.
    The reference for the zero  energy is arbitrary chosen.
  }
  \label{fig:F_V}
\end{figure}

The thermodynamics at finite temperature requires the calculation of the entropic contribution.
From the experimental point of view, the entropy is dominant over the internal energy \cite{Allen1982,Amadon2006,Decremps2011} and the necessity of computing the entropy to describe the transition at finite temperature was highlighted \cite{Amadon2006}.
The entropy, however, is made of two physical contributions, coming respectively from lattice and electrons.
It was shown both from experimental phonon spectra~\cite{Krisch2011} and ultrasonic measurements~\cite{Decremps2009} that the variation of electronic contribution is the dominant one and represents from 78\% to 85\% of the total entropy variation across the transition. 
We thus focus on the electronic contribution to entropy. 
We use the recent coupling constant integration approach developed in Ref.~\onlinecite{Pourovskii2013} to compute the free energy  

\begin{equation}
 F(U)=F(0) + \int_0^U \frac{E_{\rm int}[U']}{U'} dU',
  \label{eq:free_E}
\end{equation}
where $E_{\rm int}=\langle H_U \rangle - E_\text{DC}$. $E_{\rm int}$ is the only part of the energy which has an explicit dependence of $U$, $\langle H_U \rangle$ is computed inside the CT-Hyb \cite{Gull2011} and 
$F(0)$ is the LDA free energy.
Afterwards, we deduce the entropy from the knowledge of $F$ and $E$.

The free energy curves (solid bold lines) are plotted on the middle panel of Fig.~\ref{fig:F_V}.
As discussed above and physically expected, the entropy is weak at small volumes.
Consequently, the agreement for small volumes between internal energy which is computed directly with Eq.~\eqref{eq:energy} and free energy which is computed with the thermodynamical integration given in Eq.~\eqref{eq:free_E}, validates our approach.
Secondly, we notice that the entropy contribution ($-TS$) is very tiny for the study at 100 K. 
It shows that at lower temperatures the entropic contribution can be neglected. 
The regularity of the free energy curves suggest no negative curvature, therefore we conclude that there are no transition at low temperature as described by our DFT+DMFT scheme without \textsc{soc}.
With our accurate DFT+DMFT framework, we confirm \citet{Lanata2013}'s results obtained with the Gutzwiller approximation: neither a transition nor a clear softening of the bulk modulus are observed in the zero temperature limit without \textsc{soc}.
\footnote{In fact, a very weak softening is observed in our $P(V)$ curves  around $V=37$ \AA$^3$. 
  The same softening is surprisingly observed in the curves of \citet{Lanata2013} without \textsc{soc}, even if the authors do not mention it. 
  Anyway, we confirm the overall behavior of their work.
}
We will see in the following that the same conclusion does not apply to finite temperature.

In contrast, our conclusion does not match the results obtained by \citet{Casadei2012} who are able to describe the phase transition in cerium at zero temperature without \textsc{soc} in their DFT scheme using exact exchange and a RPA correlation energy.

We now discuss about the physics at higher temperature where the entropic contribution is important.
Considering the free energy curves on the middle panel of Fig.~\ref{fig:F_V}, we observe three characteristics.

Firstly, we note a large increase of the entropy from small volumes to large volumes.
As described in the MT and KVC models, the only existence of local moments in the $\gamma$ phase explains the important variation of entropy as presented on the lower panel of Fig.~\ref{fig:F_V}.
A detailed interpretation for this large variation is discussed in detail in the appendix~\ref{ap:entropy}.
Moreover, the critical volume defined as the volume above which entropy increases, increases when the temperature is decreased: this is just because the Kondo temperature is an increasing function of the hybridization.

Secondly, for several finite temperatures, we plot both the free energy curves $F(V)$, and the pressure $P(V)$ versus volume (solid lines) on the upper panel of Fig.~\ref{fig:F_V}.
None of those free energy curves show a negative curvature, nor a cancellation of the bulk modulus.
In other words, a first order phase transition is not observed within our scheme. 
It is hence in contradiction with experiments,  although for all the temperatures plotted except 100 K, a decrease of the slope of the pressure is clearly visible.
This implies a softening of the bulk modulus for those temperatures and reflects a crossover.
That is to say, the system is actually near a phase transition. 
The comparison of our $P(V)$ curves with experiment shows the improvement brought by the inclusion of entropy (solid lines) versus the pressure computed from the internal energy (dashed lines).
The high pressures are obviously better described with entropy.

Thirdly, from our $P(V)$ curves, one can also estimate the transitional temperature for a fixed pressure.
We clearly see that for zero pressure and a temperature from 400K to 600K, there is a transition from the $\alpha$ phase to the $\gamma$ phase.
In other words, the minimum of those free energy curves with respect to the temperature is shifted from the $\alpha$ volume to the $\gamma$ volume when the temperature is increased as experimentally expected.
For instance, at 400 K and zero pressure our free energy describes the $\alpha$ phase as stable, whereas the $\gamma$ phase becomes the most stable at 800 K.
Nonetheless, according to experiment, this transition should appear at a much lower temperature.

\section{Influence of spin-orbit coupling\label{sec:soc}}
\begin{figure}[btp]
  \centering
  \includegraphics{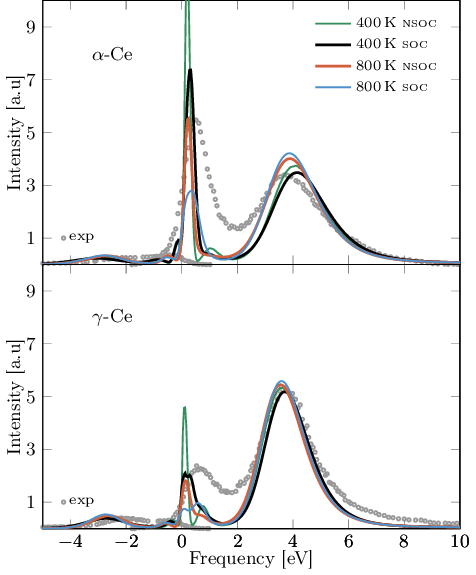}
  \caption{(color online) Spectral functions of $4f$ electrons from DFT+DMFT (solid lines) at 400 K and 800 K with and without \textsc{soc} compared to $4f$ spectra obtained from Resonant Photoemission \cite{Weschke1991} and Resonant Inverse Photoemission Spectra \cite{Grioni1997} at 300 K (dots).
    All these curves are in good agreement : a smaller quasi-particle peak for the (localized) $\gamma$ phase (lower panel) and a larger quasi-particle peak for the (more delocalized) $\alpha$ phase (upper panel). The influence of spin-orbit coupling seems to be a renormalization of the temperature as the curves at 800 K without soc and 400 K with soc are very close.
  }
  \label{fig:spectra}
\end{figure}
So far we did not include \textsc{soc}.
We argue that the lack of \textsc{soc} is equivalent to a renormalization of temperature.
Its inclusion, besides being computationally more expensive due to the appearance of the sign problem in the CT-Hyb, would lead in the atomic limit to a degeneracy of the $f$ orbitals of 6 and so to an entropy of $\ln(6)$ instead $\ln(14)$. 
Our calculations show that this atomic limit is recovered between the $\gamma$ and $\alpha$ phases.
In the degenerate Hubbard model\cite{Florens2002}, this lowest degeneracy leads to a reduced critical temperature.
So, in our study without \textsc{soc}, we can roughly expect an increase of the critical temperature. 
This argumentation stands especially for the $\gamma$ phase while it is less valid for the $\alpha$ phase since the crystal field splitting is as important as the \textsc{soc}.

For this reason, we calculated $f$ spectral functions using an analytic continuation by Maximum Entropy of imaginary time Green Function, at 400 K and 800 K, with and without spin-orbit coupling (\textsc{soc}) \footnote{With the value $U=6$ eV, we assume that the off diagonal terms of the green functions in the $J m_J$ basis are negligible}.
We aim at checking weather our assumption of renormalized temperatures is roughly valid for both phases.
They are compared on Fig.~\ref{fig:spectra} to resonant photoemission spectra \cite{Weschke1991,Grioni1997} which isolate the $f$ contribution.

In the first place, the qualitative features are all present in our scheme as in previous works \cite{Zolfl2001,Held2001,Amadon2006}.
The more localized $f$ electrons in the $\gamma$ phase produce a smaller quasi-particle peak in the $\gamma$ phase than in the $\alpha$ phase in agreement with previous studies \cite{Held2001,McMahan2003,Amadon2006}.
Moreover, the combination of the PAW scheme and the self-consistency leads to a much better description of the position of Hubbard bands compared to those works.  
We now comment on the impact of spin orbit coupling.
It appears that calculations at 400 K  with \textsc{soc} and at 800 K without \textsc{soc} are alike.
Indeed, one can check that without \textsc{soc}, 400 K is below the Kondo temperature for both phases (both have a huge quasiparticle peak) and with \textsc{soc}, 800 K is too much above the Kondo temperature (quasiparticle peak too small for both phases).
So the \textsc{soc} qualitatively gives rise to the same physics as without \textsc{soc} but at a different temperature.
We can thus expect that with \textsc{soc} for the same calculations presented here, the physical features should qualitatively be the same.

In the light of this effect, the thermodynamical quantities are compared in Tab.~\ref{tab:tab1} in which we use our result without \textsc{soc} at 800 K with experimental data at 400 K. 

\begin{figure}[tbp]
  \centering
  \includegraphics{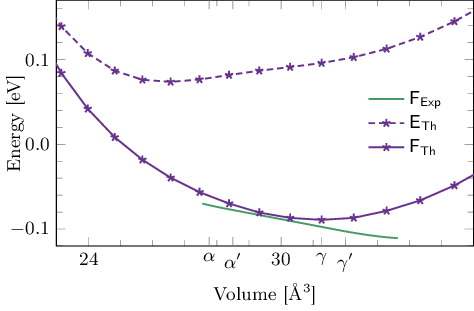}
  \caption{(color online) Theoretical internal and free energies at 800 K with respect to experimental free energy at 400 K~\cite{Decremps2011}.
  $\alpha$ and $\gamma$ indicate respectively the volume of each phase with our equilibrium volume with the same volume difference as experiments at 400 K.
  $\alpha'$ and $\gamma'$ indicate the experimental volumes at 400 K.
}
  \label{fig:zoom}
\end{figure}

\begin{table}[tbp]
  \begin{tabular}{cccc}
    \hline
    \hline
    (meV) & Theo & Theo' &  Exp \\
    \hline
    $\Delta E_{\gamma\alpha}$ & 19 & 21 & 17 \\
    $\Delta S_{\gamma\alpha}$ & 52 & 38 & 41 \\
    $\Delta F_{\gamma\alpha}$ & -33 & -17 & -24 \\
    \hline
    \hline
  \end{tabular}
  \caption{Variations of thermodynamical quantities [see Fig.~\ref{fig:zoom}] across the transition at 400 K (experiments) and 800 K (theory) at different volumes : 
  $\alpha$ and $\gamma$ indicate respectively the volume of each phase with our equilibrium volume with the same volume difference as experiments at 400 K.
  $\alpha'$ and $\gamma'$ indicate the experimental volumes at 400 K.
  The experimental electronic entropy is obtained by subtracting the lattice entropy~\cite{Decremps2011}.
}
  \label{tab:tab1}
\end{table}

First of all, the main effect can be seen qualitatively, having a look at the 600 K or 800 K free energy
curves (Fig.~\ref{fig:F_V}):  the entropic contribution inverts the stabilization of the $\alpha$ and $\gamma$ phases as shown in Ref.~\cite{Amadon2006,Decremps2011}.
Coming back to our data of  Fig.~\ref{fig:zoom}, we find out that our variation of thermodynamical quantities at 800 K are coherent with the experimental data at 400 K, as described in Tab.~\ref{tab:tab1}.
This can be nuanced according to the volume chosen for the $\alpha$ and the $\gamma$ phase.
We present two sets of volumes.
The first one, ``Theo'' stands for volumes with our equilibrium volume for the $\gamma$ phase and the $\alpha$ volume is chosen so that $\Delta V$ is the experimental one.
The second one, ``Theo' '' stands for the experimental volumes with the same $\Delta V$ at 400 K.
Even if we cannot expect a perfect agreement without \textsc{soc}, the agreement between thermodynamical data is surprisingly good.
Moreover, the Bulk modulus extracted from the 800K free energy is reduced by the entropic contribution from 36 GPa to 23 GPa, in good agreement with the experimental value of 20 GPa~\cite{Decremps2009}.

\section{Conclusion}
We carried out accurate internal and free energy calculations for the isostructural transition in cerium. 
We found that without spin-orbit coupling, no transition appears at zero temperature. 
However, above zero temperature, a crossover is observed both in temperature and pressure between the $\alpha$ and $\gamma$ phases. 
This is distinctly visible as a softening of the bulk modulus. 
We showed the leading role of entropy for the description of the equation of state of cerium. 
Finally, we discussed the role of spin-coupling coupling, and we argued that neglecting the spin-orbit coupling is roughly equivalent to a renormalization of temperature. 
Indeed both our variation of thermodynamical quantities and our spectral functions describe the experimental data at 800 K.
This establishes firmly the DFT+DMFT results for the transition and opens the way for more general scheme, including other important physical effects such as the spin-orbit coupling for the energy, the inclusion of the variation of $U$ as a function of volume \cite{McMahan1998,Aryasetiawan2006,Nilsson2013,Amadon2014} and the lattice contribution of entropy.

\acknowledgments
We thank M. Alouani, S. Biermann, L. Colombet, C. Denoual, M. Ferrero, E. Gull, F. Jollet, F. Lechermann, A.I. Lichtenstein, C. Martins and M. Torrent for discussions about this work.
We acknowledge Partnership for Advanced Computing in Europe (PRACE) for awarding us access to resource Curie based in France at Tr\'es Grand Centre de Calcul (TGCC) (Preparatory access) and to resource Marenostrum III based in Spain at Barcelona Supercomputing Center (BSC) (Regular access). 

\appendix
\section{Decomposition of the entropic contribution\label{ap:entropy}}
\begin{figure}[b] \includegraphics{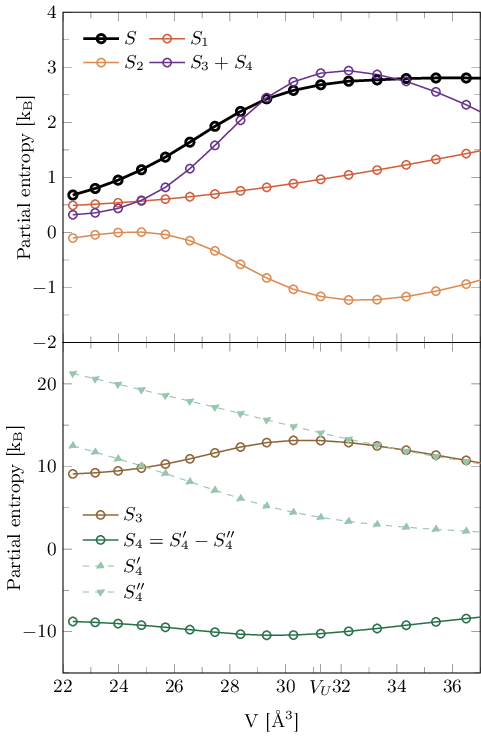}
  \caption{Different terms of the entropy as expressed in Eq.~\ref{eq:Sdecomp} computed with LDA+DMFT(CT-Hyb) at 800~K. }
  \label{fig:Sdecomp}
\end{figure}
We discuss in the following the calculation of electronic entropy.
We use the recent coupling constant integration approach used by~\citet{Pourovskii2013}. 
From the self-consistent density $\rho$, the internal energy can be computed\cite{Amadon2006} as
$E(U) = E_{LDA}(\rho) - \sum_\lambda \epsilon_\lambda^{\text{KS}} + \langle H_{\text{KS}} \rangle + E_{\rm int}[U]$
 where $E_{\rm int}[U]=\langle H_U \rangle - E_\text{DC}$. $E_{\rm int}[U]$ is the only part of the energy which has an explicit dependence on U.
$\langle H_U \rangle$ is computed inside the CT-Hyb QMC\cite{Gull2011}  and we use for $E_\text{DC}$ the double counting energy \cite{Held2001,Amadon2006,Streltsov2012}.

The coupling constant integration for the free energy reads\cite{Pourovskii2013}:
\begin{equation}
 F(U)=F(0) + \int_0^U \frac{E_{\rm int}[U']}{U'} dU'.
  \label{eq:F}
\end{equation}
From the definition of $F$, we deduce the entropy:
\begin{equation*}
 S(U)= S_{\rm LDA}(\rho_{\rm LDA})+\frac{E(U)-E(0)}{T} -\frac{1}{T}\int_0^U \frac{E_{\rm int}[U']}{U'} dU'
\end{equation*}
From these equations, we have a practical way of computing $S(U)$:
For each volume and temperature, we performed DFT+DMFT calculations for values of U from 0 eV to
6 eV. The variation of $E_{\rm int}/U$ as a function of U was splined and integrated following previous equation. We checked the convergence of the entropy  with respect to the number of values of U.
The LDA and LDA+DMFT entropies are compared on the upper part of Fig.~\ref{fig:Sdecomp}:
Firstly, the LDA+DMFT entropy is twice larger than the LDA entropy for large volumes and is coherent with the expected value of $\ln(14)$ in qualitative agreement with \cite{McMahan2003}. 
We latter discuss the modification that \textsc{soc} would bring.
Secondly, whereas in LDA, the increase is slow, in LDA+DMFT the variation is fast, and occurs in the domain of the experimental
volume of transition, when the Kondo stabilization energy disappears as the volume increases\cite{Amadon2006}.
It is thus coherent with the physical picture of the localization of electron at the transition and the exponential character of the Kondo scale as a function of hybridization.
Thirdly, as temperature decreases, the volume domain for the fast variation of $S$ is shifted upwards coherently 
with the increase of the critical volume for localization.
Finally, we note that $-TS$ contains an important negative curvature, at 400 K for example, between 25 \AA$^3$ and 30 \AA$^3$. 
It could thus contribute to the appearance of a negative curvature also in the free energy.

In order to understand more deeply the origin of the variation of $S$, we
use the expression of internal energy to decompose the DFT+DMFT entropy as:
\begin{eqnarray}
  \label{eq:Sdecomp}
 S(U)&=& \underbrace{S_{\rm LDA}}_{S_1} 
 +\underbrace{(E_{\rm LDA}(\rho)-E_{\rm LDA}(\rho_{\rm LDA}))/T}_{S_2}\\
 \nonumber  
 &&{}+ \underbrace{\frac{\langle H_{\text{KS}}\rangle-\sum_\lambda \epsilon_\lambda^{\text{KS}} }{T}}_{S_3} +\underbrace{\frac{E_{\rm int}[U]}{T}}_{S_4'}-\underbrace{\frac{\int_0^U \frac{E_{\rm int}[U']}{U'} dU'}{T}}_{S_4''}
\end{eqnarray}

We now discuss the variations of these terms as plotted in Fig. \ref{fig:Sdecomp}.
Firstly, the LDA entropy of the $f$ shell is expected to
vary from 0 to $14\ln{14}-13\ln{13}\simeq 3.60 k_{B}^{-1}$
as the dispersion of bands increases under compression.

Secondly, we can gather all the DFT+DMFT correction to the LDA entropy in two terms, namely $S_2$ and $S_3+S_4$ that are
represented on the upper part of Fig. \ref{fig:Sdecomp}.
These two terms have opposite behavior, but the most important variation comes from $S_3+S_4$.
$S_2$ is proportional to the difference between LDA energies for LDA+DMFT and LDA densities. 
It is a term which comes from the stabilization of the internal energy of the $\gamma$ phase by the DMFT density\cite{Amadon2012}.
In order to understand it, we focus on the lower part of Fig. \ref{fig:Sdecomp}, where  $S_3$, $S_4'$ and $S_4''$ are plotted.
From Eq. \ref{eq:Sdecomp}, $S_3$ is proportional to the difference of band energies. Thus, as discussed in Ref.~\cite{Amadon2006},
it first increases and thus decreases after a critical volume $V_U$ (see Fig. \ref{fig:Sdecomp}) above which the hybridization is weak enough to trigger the localization of electrons.
It thus contributes directly to the increase of entropy in DFT+DMFT near the volume of transition.
Concerning  $S_4=S_4'-S_4''$:
$S_4$ is a negative quantity, because $E_{\rm int}/(TU)$ is a decreasing function of U and thus $S_4'=1/T \int_0^U E_{\rm int}[U]/U dU'$ is  always larger than  $S_4''=1/T\int_0^U E_{\rm int}[U']/U' dU'=\int_0^U S_4'[U']/U' dU'$.
Moreover  $ E_{\rm int}/(TU)=S_4'/U$  is also a decreasing function of $V$, and 
its slope reduces largely around the U-dependent critical volume $V_U$.
$V_U$ is a decreasing function of U. Consequently, the slope of $S_4''$, which is an average of the slopes of $S_4'$  for different U, is lower --- because for small values of U, the initial slope of $S_4'$ is much reduced.
The second consequence is that above $V_U$, $S_4''$ is still a decreasing function whereas $S_4'$ is flat.
It results that $S_4$ first decreases and, above $V_U$, increases. It thus contributes directly to the increase of $S$ for large volume.
Because of this and the amplitude of $S_3$ and $S_4$, one find that $S_3+S_4$ is an increasing function of $V$ in agreement with the simple physical picture described above.

\end{document}